\documentclass{aastex701}

\usepackage{subcaption}
\usepackage[colorlinks=true]{hyperref} 

\begin{document}

\title{Hunting for Compact Object Binaries from eRASS1 Optical Counterparts through ZTF Time-domain Photometry and Multi-wavelength Surveys}

\author[0009-0001-6886-8397]{XIN-YU FANG}
\affiliation{Department of Astronomy, Xiamen University, Xiamen, Fujian 361005, People's Republic of China}
\email{fangxinyu@stu.xmu.edu.cn}  

\author[0000-0002-2912-095X]{HAO-BIN LIU}
\affiliation{Department of Astronomy, Xiamen University, Xiamen, Fujian 361005, People's Republic of China}
\email{liuhaobin@stu.xmu.edu.cn}

\author[0000-0003-3137-1851]{WEI-MIN GU}
\affiliation{Department of Astronomy, Xiamen University, Xiamen, Fujian 361005, People's Republic of China}
\email[show]{guwm@xmu.edu.cn}

\begin{abstract}

Capitalizing on the eRASS1 optical counterpart catalog, we conduct a systematic search for compact object binary (COB) candidates, with a primary focus on X-ray binaries (XRBs), by integrating ZTF time-domain photometry with multi-wavelength data. This framework establishes two independent pipelines, yielding two samples. The first sample consists of 151 periodically variable sources, from which a highly refined subset of 43 high-priority COB candidates is identified. The second sample comprises 1,958 distance-constrained sources selected on the basis of their elevated X-ray luminosities or high $\log(F_{\mathrm{X}}/F_{\mathrm{opt}})$ values. Crucially, cross-matching both samples with radio catalogs reveals seven radio-emitting sources, highlighting four promising XRB candidates. Our results underscore that coupling eROSITA with wide-field time-domain photometric and multi-wavelength surveys offers a highly efficient strategy for uncovering the hidden population of COBs. 

\end{abstract}


\section{INTRODUCTION} 

Compact object binaries (COBs) are binary systems consisting of a white dwarf (WD), neutron star (NS), or black hole (BH), together with a companion star. Investigations into these systems have provided crucial understanding of accretion physics, the equation of state of dense matter, and the late-stage evolution of massive stars \citep{Frank1992,BURGIO2021,Belloni2023}. Despite the steadily increasing number of known systems, the current census of COBs,  especially X-ray binaries (XRBs), is still far from comprehensive. In terms of BHs, theoretical models predict that the Milky Way hosts more than $10^8$ BHs, yet only about 20 have been dynamically confirmed so far \citep{Corral-Santana2016,Olejak2020}.

X-ray surveys have provided an important avenue for discovering a diverse population of intriguing COBs, particularly XRBs, since the ROSAT All-Sky Survey \citep{TRUMPER1982,Boller2016}. Subsequent observatories such as the XMM-Newton, the Chandra X-ray Observatory, and the Swift Observatory have provided improved sensitivity and spatial resolution. Nevertheless, because the primary scientific designs of these missions favor deep and localized observations over wide-field surveys, they lack global all-sky coverage \citep{Jonker2011,Bahramian2021}. The advent of the first eROSITA All-Sky Survey (eRASS1), delivering uniform X-ray coverage of the entire sky, has already cataloged over 900,000 sources in the western Galactic hemisphere  \citep{Predehl2021,Merloni2024}. The increased population of eRASS1 X-ray sources provides a massive reservoir for discovering new COB candidates through cross-matching X-ray and optical catalogs \citep{Rodriguez2025b,Muoz-Giraldo2026}. 

XRBs predominantly reside in a quiescent state, unlike systems in outburst ($L_{\mathrm{X}}$ $\sim 10^{35}\text{-}10^{38}$ erg s$^{-1}$) that are readily detected in these X-ray surveys \citep{Remillard2006,Plotkin2013}. The quiescent XRBs have $L_{\mathrm{X}} \approx 10^{30}\text{-}10^{34}$ erg s$^{-1}$,  overlapping with those of cataclysmic variables (CVs) and active binaries (ABs) \citep{Heinke2005,Reis2013}. Hence, revealing quiescent XRBs from these contaminations requires extra characterization through follow-up multi-wavelength identification programs. An effective approach is the empirical cut line between $F_{\mathrm{X}}/F_{\mathrm{opt}}$ and the $G_{\mathrm{BP}} - G_{\mathrm{RP}}$ proposed by \cite{Rodriguez2024} to separate active stars or ABs from accreting compact objects. This method has been successfully applied in combination with complementary datasets to identify COBs. For example, \cite{Zhao2026} conducted a systematic search by integrating Gaia astrometry with $F_{\mathrm{X}}/F_{\mathrm{opt}}$ diagnostics, revealing a population of COB candidates potentially accelerated by supernova natal kicks. Nevertheless, only a small fraction of eRASS1 sources have been classified through detailed multi-wavelength studies, suggesting that many potential COBs remain unidentified \citep{Maan2025,Wang2026}.

A primary challenge in identifying these systems is the lack of reliable optical counterparts, a difficulty that is particularly pronounced for eRASS1 sources due to their relatively modest spatial resolution \citep{Predehl2021}. Recently, using the Bayesian NWAY algorithm with trained priors, \cite{Salvato2025} performed large-scale probabilistic cross-matching within the Legacy Survey DR10 (LS10), identifying candidate optical counterparts for 656,614 eRASS1 sources. Although their primary aim was to construct complete AGN samples, their work also provides well-defined optical counterpart catalogs for eRASS1 sources, thereby facilitating and accelerating the identification of COBs. With reliable optical counterparts available, subsequent time-domain photometric analyses and multi-wavelength investigations can provide powerful tools for identifying COB candidates. In particular, many COBs exhibit periodic optical variability, including ellipsoidal modulation, eclipses, and irradiation effects. Meanwhile, COBs, especially XRBs, display a broad range of observational characteristics across multiple wavebands, making detailed multi-wavelength studies essential for constraining their physical nature and population properties  \citep{Russell2006, Xu2026}.

In this study, we utilize the eRASS1 counterpart catalog built by \cite{Salvato2025} and present a comprehensive search of COB candidates especially XRB candidates by integrating optical time-domain survey with multi-wavelength archival data. This paper is organized as follows. In Section \ref{sec2}, we introduce the eRASS1 optical counterpart catalog and the selection processes for our two samples, which are primarily based on optical periodic variations and X-ray properties, respectively. In Section \ref{sec3}, we analyze the characteristics of these two samples by cross-matching them with multi-wavelength data. Finally, we summarize our findings and outline our main conclusions in Section \ref{sec4}.

\section{Sample creation} \label{sec2}
\subsection{eRASS1 Optical Counterpart Catalogs}

\cite{Salvato2025} provide three catalogs of optical/infrared counterparts to eRASS1 X-ray point sources, identified using Gaia DR3, CatWISE2020, and LS10. Among these, the LS10 counterparts are considered the most reliable, owing to the greater depth and the larger number of features such as morphology incorporated in the matching procedure. We therefore adopt the LS10 catalog as our primary source of optical counterparts to eRASS1.

The parameter \texttt{class\_gal\_exgal} provides a probabilistic classification of the counterparts, where negative values indicate stellar sources and positive values indicate extragalactic objects. To initially identify Galactic accreting compact object candidates, we retain only the best optical counterparts (i.e., \texttt{class\_gal\_exgal} $<$ 0 and \texttt{NWAY\_match\_flag} = 1) that lie above the “X-ray main sequence" defined by \cite{Rodriguez2024}. This selection yields a sample of 13,631 sources.

\subsection{First Sample: Optically Variable Sources with Periodic Modulation}
\begin{figure*}[ht!]
\includegraphics[width=1.0\textwidth]{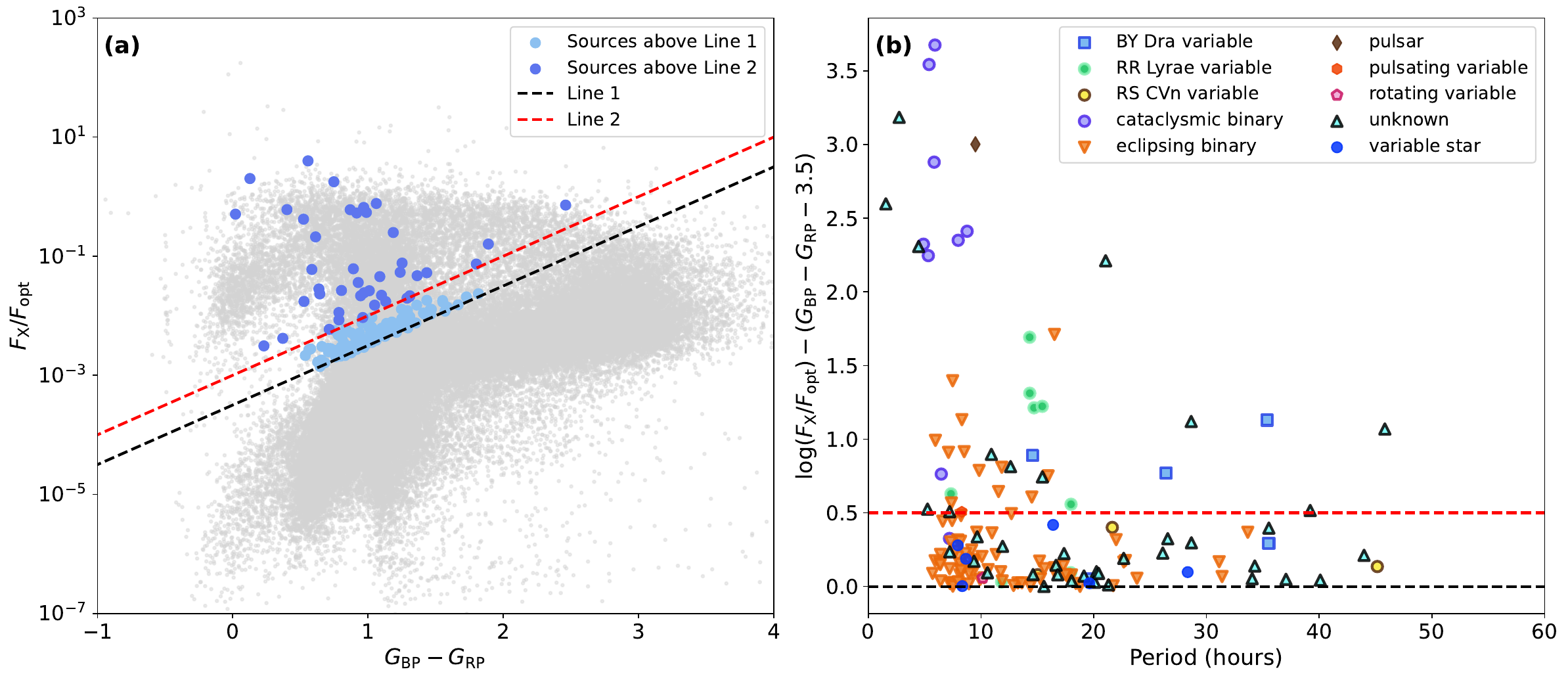} 
\caption{Panel (a): $F_{\mathrm{X}}/F_{\mathrm{opt}}$ vs $G_{\mathrm{BP}} - G_{\mathrm{RP}}$ color plotted for the sources in first sample. Sky blue points are sources above Line 1 and navy blue points are sources above Line 2. Line 1 denote $\log(F_{\mathrm{X}}/F_{\mathrm{opt}}) = (G_{\mathrm{BP}} - G_{\mathrm{RP}}) - 3.5$, while Line 2 denote $\log(F_{\mathrm{X}}/F_{\mathrm{opt}}) = (G_{\mathrm{BP}} - G_{\mathrm{RP}}) - 3$. Panel  (b): Distribution of the offset from the “X-ray main sequence” for all 151 sources vs photometric period. The y-axis shows the vertical deviation from the empirical relation $\log(F_{\mathrm{X}}/F_{\mathrm{opt}}) = (G_{\mathrm{BP}} - G_{\mathrm{RP}}) - 3.5$ (black dashed line), and the x-axis represents the photometric period. Markers of different colors and shapes indicate sources with classifications retrieved from SIMBAD.
\label{F1}}
\end{figure*}

An essential aspect to consider is residual extragalactic contaminants.  Active galactic nucleis (AGNs) are strong X-ray emitters that can also satisfy the high $F_{\mathrm{X}}/F_{\mathrm{opt}}$ criterion, making it difficult to distinguish COBs from AGNs using the “X-ray main sequence” selection alone \citep{Schwope2024}. A widely adopted approach to mitigate this contamination is to exploit variability properties  \citep{Liu2025,Wang2025,Galiullin2024}. COBs often exhibit short-term periodic photometric variability associated with orbital motion, particularly ellipsoidal modulation of the companion star \citep{Wang2025}. While AGNs usually show the long-term variability from days to years \citep{Ulrich2003,smith2006}. 
Following this strategy, we cross-match the selected eRASS1/LS10 sources with objects from the ZTF DR21 catalog and search for periodic modulation in both the g and r band light curves. Periodic variability is identified using the Lomb-Scargle periodogram, with a searched period range of 0.01-10 days. For each source, we further visually inspect the phase-folded light curve at the period corresponding to the maximum power. This procedure yields an initial sample of 151 sources exhibiting relatively clear periodic variability. Based on the “X-ray main sequence”, we also adopt a modified Line 2 to select out source with more likelihood to be COBs:   
$\log(F_{\mathrm{X}}/F_{\mathrm{opt}}) = (G_{\mathrm{BP}} - G_{\mathrm{RP}}) - 3$. The optical flux is estimated as $F_{\rm opt} = 10^{-0.4 (G - 4.83)} L_\odot / (4\pi (10~{\rm pc})^2)$, while X-ray flux is derived directly from the eROSITA \texttt{ML\_FLUX\_1} in the 0.2-2.3 keV band.
Figure \ref{F1} (a) plots $F_{\mathrm{X}}/F_{\mathrm{opt}}$ vs $G_{\mathrm{BP}} - G_{\mathrm{RP}}$ for the selected eRASS1 sources with a detection of the periodic modulation in ZTF data. We find that those most periodic sources cluster around the Line 1 and 43 sources are located above the Line 2 suggesting they are promising COBs candidates. 
The SIMBAD crossing-matching results show that 9 sources are categorized as CV, one source is classified as pulsar. Additionally, 78 sources are identified as eclipsing binaries, while 25 are categorized as other types of variable stars (including BY Dra, pulsating, rotating, RR Lyrae, and RS CVn variables). Notably, although 1eRASS J052316.7-252738 is cataloged as an eclipsing binary in SIMBAD, several studies argue that it is a redback. The remaining 38 sources currently lack definitive classifications. Figure \ref{F1} (b)  presents the distribution of SIMBAD classifications for the 151 sources in our first sample. Table \ref{tab1}  summarizes the properties of the 43 sources which are above Line 2 in the first sample.

\begin{deluxetable*}{lccccccccc}
\tabletypesize{\scriptsize}
\tablecaption{Summary of the 43 periodic sources located above Line 2 in the first sample (The full catalog of 151 sources is available in a machine-readable format). \label{tab1}} 
\tablewidth{0pt}
\tablehead{
\colhead{eROSITA Name} & \colhead{R.A.} & \colhead{Decl.} & \colhead{$P_{\mathrm{ph}}$ (day)} &  \colhead{$F_{\mathrm{X}}$ ($\rm erg\,s^{-1}\,cm^{-2}$)} & \colhead{$G$} & \colhead{$G_{\mathrm{BP}} - G_{\mathrm{RP}}$} & \colhead{$F_{\mathrm{X}}/F_{\mathrm{opt}}$} & \colhead{Type}\\
\colhead{(1)} & \colhead{(2)} & \colhead{(3)} & \colhead{(4)} & \colhead{(5)} & \colhead{(6)} & \colhead{(7)} & \colhead{(8)} & \colhead{(9)}
}
\startdata
1eRASS J030951.2-183053 & 47.463 & -18.515 & 0.346 & $2.405 \times 10^{-13}$ & 14.958 & 1.289 & 0.020 & pulsating variable \\
1eRASS J034220.8-102038 & 55.586 & -10.346 & 0.494 & $3.039 \times 10^{-14}$ & 18.146 & 1.363 & 0.047 & eclipsing binary \\
1eRASS J041920.0+072546 & 64.834 & 7.429 & 0.607 & $3.359 \times 10^{-13}$ & 14.695 & 0.946 & 0.022 & BY Dra variable \\
1eRASS J051248.0-055001 & 78.201 & -5.835 & 1.194 & $5.550 \times 10^{-14}$ & 16.872 & 0.804 & 0.027 & unknown \\
1eRASS J051541.4+010440 & 78.923 & 1.078 & 0.333 & $3.942 \times 10^{-12}$ & 15.730 & 0.968 & 0.659 & cataclysmic binary \\
1eRASS J051723.1-115400 & 79.348 & -11.900 & 0.306 & $2.762 \times 10^{-14}$ & 17.014 & 1.049 & 0.015 & RR Lyrae variable \\
1eRASS J053623.9-113343 & 84.099 & -11.561 & 0.312 & $3.250 \times 10^{-14}$ & 18.363 & 0.892 & 0.061 & eclipsing binary \\
1eRASS J054940.1-175449 & 87.415 & -17.916 & 0.355 & $4.825 \times 10^{-14}$ & 16.927 & 0.971 & 0.024 & eclipsing binary \\
1eRASS J063532.7-225751 & 98.889 & -22.962 & 0.665 & $5.718 \times 10^{-14}$ & 14.831 & 0.370 & 0.004 & eclipsing binary \\
1eRASS J072128.1+313958 & 110.366 & 31.662 & 1.910 & $5.137 \times 10^{-14}$ & 17.537 & 1.088 & 0.045 & unknown \\
1eRASS J074406.9+081701 & 116.030 & 8.281 & 0.301 & $6.023 \times 10^{-14}$ & 15.647 & 0.962 & 0.009 & unknown \\
1eRASS J081210.2+040352 & 123.043 & 4.064 & 0.225 & $1.564 \times 10^{-12}$ & 18.689 & 0.557 & 3.986 & cataclysmic binary \\
1eRASS J081352.2+281318 & 123.467 & 28.221 & 0.244 & $3.495 \times 10^{-13}$ & 18.269 & 0.401 & 0.605 & cataclysmic binary \\
1eRASS J084303.1-014859 & 130.765 & -1.816 & 0.346 & $6.844 \times 10^{-14}$ & 17.795 & 1.252 & 0.077 & eclipsing binary \\
1eRASS J084617.0+245345 & 131.571 & 24.896 & 0.365 & $2.219 \times 10^{-13}$ & 18.758 & 0.869 & 0.603 & cataclysmic binary \\
1eRASS J085630.6+210651 & 134.123 & 21.115 & 0.644 & $4.857 \times 10^{-14}$ & 16.816 & 1.100 & 0.022 & unknown \\
1eRASS J091638.2-162123 & 139.162 & -16.353 & 0.526 & $3.044 \times 10^{-14}$ & 19.472 & 1.890 & 0.160 & unknown \\
1eRASS J093757.4-171014 & 144.490 & -17.170 & 0.482 & $5.773 \times 10^{-13}$ & 13.092 & 0.785 & 0.008 & eclipsing binary \\
1eRASS J094146.1+060933 & 145.444 & 6.159 & 0.410 & $2.493 \times 10^{-13}$ & 15.985 & 1.434 & 0.053 & eclipsing binary \\
1eRASS J094558.1+292252 & 146.493 & 29.381 & 0.247 & $6.029 \times 10^{-13}$ & 18.981 & 0.128 & 2.012 & cataclysmic binary \\
1eRASS J095942.5-123407 & 149.925 & -12.568 & 0.605 & $4.030 \times 10^{-14}$ & 16.754 & 1.131 & 0.017 & eclipsing binary \\
1eRASS J100937.6+080846 & 152.409 & 8.147 & 0.248 & $8.754 \times 10^{-14}$ & 17.141 & 1.238 & 0.054 & eclipsing binary \\
1eRASS J101508.8-030834 & 153.789 & -3.142 & 0.270 & $6.532 \times 10^{-14}$ & 14.368 & 0.231 & 0.003 & cataclysmic binary \\
1eRASS J101822.1-073700 & 154.589 & -7.614 & 0.597 & $5.155 \times 10^{-14}$ & 17.015 & 0.637 & 0.028 & RR Lyrae variable \\
1eRASS J122623.4+105459 & 186.600 & 10.916 & 1.474 & $6.212 \times 10^{-14}$ & 17.083 & 0.928 & 0.036 & BY Dra variable \\
1eRASS J122928.9+013749 & 187.371 & 1.634 & 0.613 & $4.455 \times 10^{-14}$ & 16.650 & 0.527 & 0.017 & RR Lyrae variable \\
1eRASS J124334.2+121109 & 190.895 & 12.187 & 0.597 & $5.006 \times 10^{-14}$ & 17.866 & 0.585 & 0.060 & RR Lyrae variable \\
1eRASS J141148.4+052520 & 212.954 & 5.420 & 0.643 & $2.908 \times 10^{-14}$ & 17.430 & 0.644 & 0.023 & RR Lyrae variable \\
1eRASS J145358.8-115401 & 223.495 & -11.897 & 0.750 & $3.690 \times 10^{-14}$ & 15.684 & 0.714 & 0.006 & RR Lyrae variable \\
1eRASS J150354.1-220711 & 225.975 & -22.120 & 0.204 & $5.355 \times 10^{-13}$ & 18.063 & 1.061 & 0.768 & cataclysmic binary \\
1eRASS J151625.3-133931 & 229.110 & -13.659 & 0.297 & $3.715 \times 10^{-14}$ & 17.297 & 1.010 & 0.026 & eclipsing binary \\
1eRASS J152610.9-102514 & 231.546 & -10.420 & 1.101 & $1.142 \times 10^{-13}$ & 15.165 & 0.785 & 0.011 & BY Dra variable \\
1eRASS J162717.8-194938 & 246.823 & -19.828 & 0.308 & $3.545 \times 10^{-14}$ & 18.467 & 1.801 & 0.074 & eclipsing binary \\
1eRASS J052316.7-252738 & 80.821 & -25.460 & 0.688 & $7.180 \times 10^{-13}$ & 16.526 & 1.187 & 0.250 & eclipsing binary \\
1eRASS J085909.0+053652 & 134.788 & 5.615 & 0.222 & $3.780 \times 10^{-13}$ & 18.065 & 0.988 & 0.543 & cataclysmic binary \\
1eRASS J102347.6+003838 & 155.949 & 0.645 & 0.396 & $6.700 \times 10^{-12}$ & 16.232 & 0.748 & 1.778 & pulsar \\
1eRASS J040423.8-270245 & 61.100 & -27.045 & 0.455 & $9.530 \times 10^{-14}$ & 19.870 & 2.461 & 0.721 & unknown \\
1eRASS J050325.1-241653 & 75.855 & -24.281 & 0.878 & $2.250 \times 10^{-13}$ & 17.602 & 0.613 & 0.211 & unknown \\
1eRASS J051229.7+040144 & 78.125 & 4.029 & 1.634 & $1.610 \times 10^{-13}$ & 15.463 & 1.307 & 0.021 & unknown \\
1eRASS J080407.3-002218 & 121.033 & -0.371 & 0.065 & $1.600 \times 10^{-13}$ & 18.715 & 0.524 & 0.418 & unknown \\
1eRASS J130541.8+180103 & 196.425 & 18.018 & 0.114 & $2.530 \times 10^{-12}$ & 15.929 & 0.021 & 0.508 & unknown \\
1eRASS J135239.1-243602 & 208.162 & -24.602 & 0.186 & $9.340 \times 10^{-14}$ & 19.559 & 0.917 & 0.531 & unknown \\
1eRASS J151037.3-264235 & 227.653 & -26.706 & 0.219 & $5.910 \times 10^{-14}$ & 16.580 & 1.309 & 0.022 & unknown \\
\enddata
\tablecomments{(1) eROSITA X-ray IAU name. (2) Right Ascension (R.A.) and (3) Declination (Decl.) of the LS10 optical counterpart from \cite{Salvato2025}. (4) Photometric period ($P_{\mathrm{ph}}$) derived from ZTF light curves. (5) eROSITA X-ray flux in the 0.2-2.3 keV band. (6) Gaia $G$-band magnitude. (7) Gaia $G_{\rm BP}-G_{\rm RP}$ color. (8) X-ray-to-optical flux ratio. (9) Object classification according to the SIMBAD database.}
\end{deluxetable*}

\subsection{Second Sample: Sources with High $L_{\mathrm{X}}$ or High $\log (F_{\mathrm{X}}/F_{\mathrm{opt}})$}
Apart from periodic optical variability, astrometric information from Gaia provides a powerful complementary constraint. In particular, the high-precision parallaxes delivered by Gaia enable direct distance measurements for a vast number of sources, offering a crucial advantage in distinguishing Galactic objects from extragalactic contaminants  \citep{Bailer-Jones2021}.

We therefore further refine our sample by requiring significant and reliable Gaia astrometric measurements, retaining only sources with well-constrained parallaxes and proper motions ($\texttt{Gaia\_PM\_SNR > 5}$ and $\texttt{Gaia\_PARALLAX\_SNR > 5}$). Distances are estimated from the parallaxes, primarily using the geometric relation $d = 1/\varpi$ and subsequently used to derive the X-ray luminosities $L_{\mathrm{X}}$ for each source. Here, $L_{\mathrm{X}}$ is computed directly from the eROSITA 0.2-2.3 keV flux ($\texttt{ML\_FLUX\_1}$) without any corrections ($L_{\mathrm{X}} = 4\pi d^2 F_{\mathrm{X}}$).

\begin{figure*}[ht!]
\centering 
\includegraphics[width=0.7\textwidth]{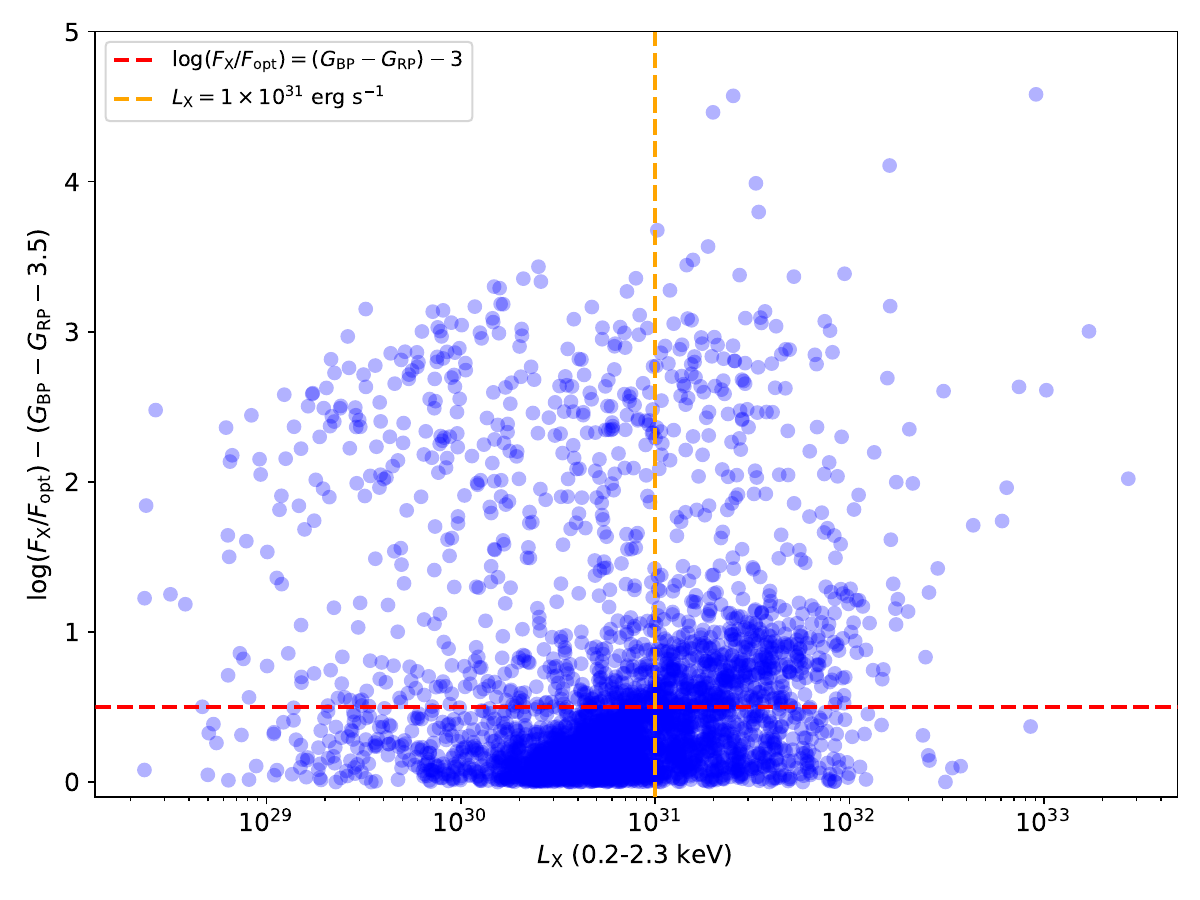} 
\caption{X-ray luminosities to $\log (F_{\mathrm{X}}/F_{\mathrm{opt}}) - (G_{\mathrm{BP}} - G_{\mathrm{RP}}-3.5) $ distribution of sources with reliable distance. The X-ray luminosities $L_{\mathrm{X}}$ is derived by combining the eROSITA flux ($\texttt{ML\_FLUX\_1}$) parameter with Gaia parallaxes. Based on this distribution, the second sample was explicitly selected from the upper-left, upper-right, and lower-right regions of the parameter space.
\label{F2}}
\end{figure*}

We then apply additional selection criteria, requiring either $L_{\mathrm{X}} > 10^{31}\,\mathrm{erg\,s^{-1}}$ or $\log (F_{\mathrm{X}}/F_{\mathrm{opt}}) - (G_{\mathrm{BP}} - G_{\mathrm{RP}}) > -3.0$, which yields a sample of 1,958 sources, as shown in Figure \ref{F2}. The first few rows of this sample are presented in Table \ref{tab2} as an example, while the full catalog is available in the electronic version. These criteria are designed to isolate relatively X-ray bright systems or sources with high $F_{\mathrm{X}}/F_{\mathrm{opt}}$ which are more promising COBs.

eROSITA's effective energy range is concentrated in the soft X-ray band (0.2-2.3 keV), which inherently limits flux characterization for harder sources \citep{Predehl2021}. To improve source classification and obtain more robust X-ray constraints, we cross-match the 1958 sources sample with other major X-ray catalogues, including the Chandra Source Catalog (CSC 2.1, \cite{Evans2024}) and the XMM-Newton Serendipitous Source Catalog (4XMM-DR14, \cite{Webb2020}). This cross-match yields 83 sources from XMM-Newton and 2 sources from Chandra (see Table \ref{tab2} for details). Among these matched sources, 46 XMM-Newton sources are identified as CVs, 3 as pulsars, and one as a low-mass X-ray binary (LMXB) candidate.

\begin{deluxetable*}{lcccccccccccc}
\tabletypesize{\scriptsize}
\tablecaption{A subset of sources in sample 2 (The full catalog of 1,958 sources is available in a machine-readable format). \label{tab2}} 
\tablewidth{0pt}
\tablehead{
\colhead{eROSITA Name} & \colhead{R.A.} & \colhead{Decl.} & \colhead{Plx (mas)} &  \colhead{D (kpc)} & \colhead{$S/N_{\mu}$} & \colhead{$S/N_{\varpi}$} & \colhead{$G$} & \colhead{$L_{\mathrm{X}}$ ($\rm erg\,s^{-1}$)}& \colhead{$d_{\rm L2}$}& \colhead{Note} & \colhead{$F_{\mathrm{X}}$} & \colhead{$F_{\mathrm{R}}$}\\
\colhead{(1)} & \colhead{(2)} & \colhead{(3)} & \colhead{(4)} & \colhead{(5)} & \colhead{(6)} & \colhead{(7)} & \colhead{(8)} & \colhead{(9)} & \colhead{(10)} & \colhead{(11)} & \colhead{(12)} & \colhead{(13)}
}
\startdata
1eRASS J000155.3-670742 & 0.480 & -67.129 & 0.712 & 1.405 & 659.707 & 35.140 & 14.848 & $6.42 \times 10^{32}$ & 1.461& - & -& - \\
1eRASS J000207.6-374920 & 0.530 & -37.821 & 2.358 & 0.424 & 968.583 & 37.016 & 17.041 & $1.65 \times 10^{30}$ & 1.110 & - & -& -\\
1eRASS J000633.5-690035 & 1.640 & -69.009 & 4.113 & 0.243 & 1678.619 & 186.540 & 15.103 & $1.05 \times 10^{31}$ & 1.587 & - & -& -\\
1eRASS J000635.9-753134 & 1.642 & -75.526 & 1.091 & 0.917 & 644.854 & 99.983 & 13.390 & $3.38 \times 10^{31}$ & 0.226 & - & -& -\\
1eRASS J000714.3-533904 & 1.810 & -53.651 & 1.883 & 0.531 & 39.181 & 7.989 & 19.360 & $8.23 \times 10^{30}$ & 2.480 & - & -& -\\
1eRASS J000743.4-695943 & 1.930 & -69.997 & 2.130 & 0.470 & 96.863 & 11.626 & 19.153 & $8.31 \times 10^{30}$ & 2.611 & - & -& -\\
1eRASS J000959.3-450513 & 2.496 & -45.086 & 0.410 & 2.438 & 123.524 & 5.624 & 17.136 & $3.52 \times 10^{31}$ & 0.632& - & -& - \\
1eRASS J001052.4-764645 & 2.721 & -76.779 & 1.180 & 0.847 & 92.096 & 8.094 & 18.534 & $4.39 \times 10^{30}$ & 0.048 & - & -& -\\
1eRASS J001110.9-745859 & 2.795 & -74.982 & 0.474 & 2.111 & 221.583 & 9.984 & 16.549 & $1.82 \times 10^{31}$ & 0.056 & - & -& -\\
1eRASS J001159.3-554155 & 2.998 & -55.698 & 2.294 & 0.436 & 82.070 & 10.469 & 19.534 & $1.82 \times 10^{30}$ & 2.161 & - &-& -\\
\enddata
\tablecomments{(1) eROSITA X-ray IAU name. (2) Right Ascension (R.A.) and (3) Declination (Decl.) of the LS10 optical counterpart \citep{Salvato2025}. (4) Gaia parallax. (5) Distance derived from the Gaia parallax. (6) \texttt{Gaia\_PM\_SNR} (7) and \texttt{Gaia\_PARALLAX\_SNR} from \citet{Salvato2025}. (8) Gaia $G$-band magnitude. (9) X-ray luminosity estimated from the parallax and eROSITA X-ray flux. (10) Distance relative to Line 2, defined as $d_{\rm L2} = \log (F_{\mathrm{X}}/F_{\mathrm{opt}}) - (G_{\mathrm{BP}} - G_{\mathrm{RP}}) + 3.0$. (11) Flag indicating the presence of auxiliary data from 4XMM-DR14, CSC 2.1, VLASS, or RACS. (12) X-ray flux ($\text{erg s}^{-1}\text{ cm}^{-2}$) from XMM-Newton (0.2-12 keV) or Chandra (0.5-7 keV), where available. (13) Radio flux density ($\text{mJy}$) from VLASS or RACS, where available. }
\end{deluxetable*}

\section{Multi-wavelength Analysis}\label{sec3}

\subsection{Investigating the Ultraviolet-to-X-ray Properties of CVs and XRBs}

X-ray-to-optical flux ratios $F_{\mathrm{X}}/F_{\mathrm{opt}}$ have long been recognized as effective diagnostics for selecting accreting compact objects  \citep{Tranin2022,Wang2026,Liu2025}. However, CVs can occupy overlapping regions in the conventional $F_{\mathrm{X}}/F_{\mathrm{opt}}$ parameter space \citep{Bao2025}. To further distinguish candidate samples, it is therefore necessary to exploit additional differences in the spectral energy distribution (SED) between CVs and XRBs. Hence, the ultraviolet-to-X-ray flux ratio $\log(F_{\mathrm{UV}}/F_{\mathrm{X}})$ may provide a potentially powerful discriminator between these populations.

We therefore explore the use of $\log(F_{\mathrm{UV}}/F_{\mathrm{X}})$ in combination with the X-ray hardness ratio (HR) as a practical proxy for separating CVs from qLMXBs. The ultraviolet flux is derived from the GALEX near-ultraviolet (NUV) magnitude as $F_{\rm UV} = 10^{-(m_{\rm NUV} - 20.08)/2.5} \times 2.06 \times 10^{-16} \times 732$, where 732  $\mathrm{\AA}$ is the effective bandwidth of the GALEX NUV band \citep{Morrissey2007}. To evaluate this diagnostic, we cross-match the CV catalog compiled by \cite{Ritter2003} with GALEX ultraviolet data. As a comparison sample, we similarly cross-match confirmed quiescent LMXBs with non-degenerate donor stars from \cite{Fortin2024} with GALEX ultraviolet data. Magnetic CVs were intentionally excluded from this analysis, as their accretion columns and shock-heated post-shock regions produce substantially harder X-ray spectra, often resembling those of LMXBs and thus potentially biasing the comparison \citep{Mukai2017}.

\begin{figure*}[ht!]
\centering  
\includegraphics[width=1\textwidth]{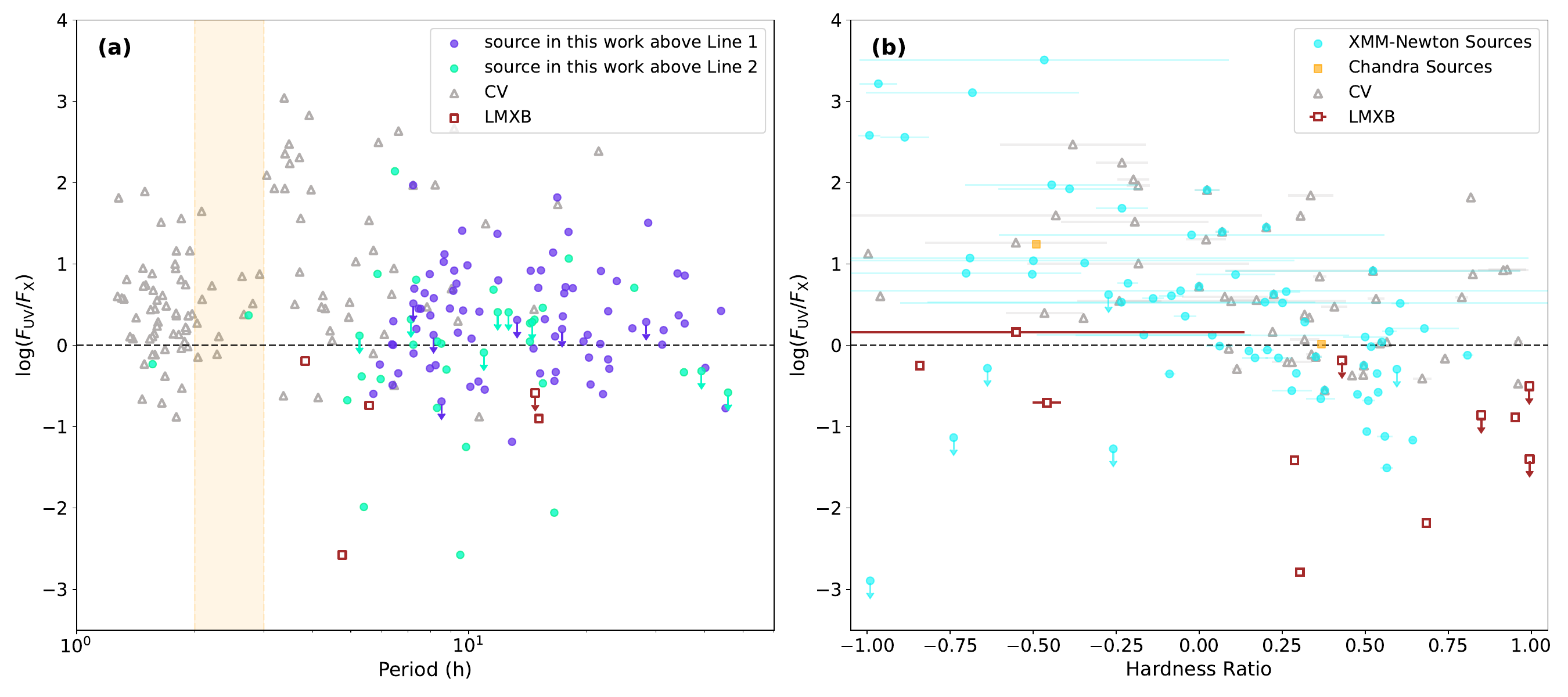} 
\caption{
Panel (a) illustrates the distribution of $\log(F_{\mathrm{UV}}/F_{\mathrm{X}})$ against photometric period for Sample 1, together with CVs from \cite{Ritter2003} and LMXBs from \cite{Fortin2024}. For all sources in this plot, the X-ray fluxes are consistently taken from the eRASS1 $\texttt{ML\_FLUX\_1}$ data. Conversely, Panel (b) displays the distribution of $\log(F_{\mathrm{UV}}/F_{\mathrm{X}})$ versus HR for Sample 2, where the X-ray measurements for the reference CVs and LMXBs are extracted from the 4XMM-DR14 catalogue.  In panel (b), light blue circles and yellow squares represent sources with XMM-Newton and Chandra counterparts, respectively. In both panels, arrows indicate NUV upper limits for sources without significant ultraviolet detections. The black dashed line at $\log(F_{\mathrm{UV}}/F_{\mathrm{X}})=0$ is shown as a reference for source classification.
\label{F3}}
\end{figure*}

As shown in Figure \ref{F3} (a), we find that a threshold at $\log(F_{\mathrm{UV}}/F_{\mathrm{X}})=0$ can serve as a reference line for broadly distinguishing these populations. Using eROSITA fluxes, about $84\%$ of non-magnetic CVs lie above this line, while all confirmed LMXBs fall below it. For comparison, in Figure \ref{F3} (b), using XMM-Newton fluxes, roughly $24\%$ of CVs lie below the line, and approximately $91\%$ of LMXBs remain below it. The X-ray HR for XMM-Newton is defined as $\mathrm{HR} = (P_4 + P_5 - P_1 - P_2 - P_3)/(P_1 + P_2 + P_3 + P_4 + P_5)$, with energy bands $P_1: 0.2-0.5$ keV, $P_2: 0.5-1.0$ keV, $P_3: 1.0-2.0$ keV, $P_4: 2.0-4.5$ keV, and $P_5: 4.5-12$ keV. The HR for Chandra sources is $\mathrm{HR} = (P_2 - P_1)/(P_1 + P_2)$, with $P_1: 0.2-1.2$ keV and $P_2: 2-7$ keV. In our comparison sample, the X-ray HR derived from XMM-Newton and Chandra data does not show a strong separation between CVs and LMXBs.

We then apply this diagnostic framework to our final candidate samples. Both Sample 1 (151 sources) and refined Sample 2 (85 sources) were cross-matched with GALEX using a $5^{\prime\prime}$ search radius. For objects located within the GALEX survey footprint but remaining undetected, NUV flux upper limits are estimated from the typical 5$\sigma$ limiting magnitudes of the corresponding GALEX surveys (AIS: $\sim20.8$ mag; MIS: $\sim22.7$ mag; DIS: $\sim24.4$ mag), yielding upper limits on $\log(F_{\mathrm{UV}}/F_{\mathrm{X}})$ \citep{Bianchi2017}.

For Sample 1, there are 38 sources that fall below $\log(F_{\mathrm{UV}}/F_{\mathrm{X}})=0$. Notably, sources located above Line 2 tend to exhibit systematically shorter periods and generally show lower $\log(F_{\mathrm{UV}}/F_{\mathrm{X}})$ values. For Sample 2, a total of 28 sources satisfy the criterion $\log(F_{\mathrm{UV}}/F_{\mathrm{X}}) < 0$, of which 18 are classified as CVs and 5 remain unclassified in SIMBAD. This indicates that $\log(F_{\mathrm{UV}}/F_{\mathrm{X}})$  has certain limitations in cleanly separating CVs and LMXBs, as a substantial number of sources may correspond to magnetic CVs. In such cases, enhanced X-ray emission and harder spectral characteristics can lead to partial overlap with the parameter space typically occupied by LMXBs.  Consequently, these five unclassified sources, regarded as potential XRB candidates, should be treated with caution, and their true nature warrants further verification.

In the two samples, eRASS1 photometry in the 2.3-5 keV band is available for 7/151 (4.6\%) and 99/1958 (5.1\%) sources, respectively. According to SIMBAD, four of the seven sources in Sample 1 have been classified as CVs or pulsars. Among the remaining three sources, 1eRASS J093757.4-171014 has a $L_{\mathrm{X}}$ (2.3-5 keV) exceeding $10^{31}\ {\rm erg\,s^{-1}}$ and may therefore warrant further investigation as a potential COB. The unclassified source 1eRASS J130541.8+180103
 has a lower hard-band luminosity of approximately $2.2\times10^{30}\ {\rm erg\,s^{-1}}$ but exhibits a relatively high $F_{\rm X}/F_{\rm opt}$, making it another potential COB candidate. However, the limited availability of eROSITA hard-band measurements, together with the sparse coverage of archival XMM-Newton and Chandra observations, prevents a comprehensive X-ray characterization of both samples. Therefore, we adopted the soft X-ray HR diagnostics based entirely on eROSITA photometry. Following the approach of \citet{Schwope2024}, who showed that soft X-ray colors can provide useful diagnostics for CVs, especially magnetic CVs, we constructed the HR1-HR2 diagram for all sources in both samples. We then investigated whether these diagnostics can provide additional constraints on the nature of the X-ray sources.

 \begin{figure}[ht!]
\centering  
\includegraphics[width=1\textwidth]{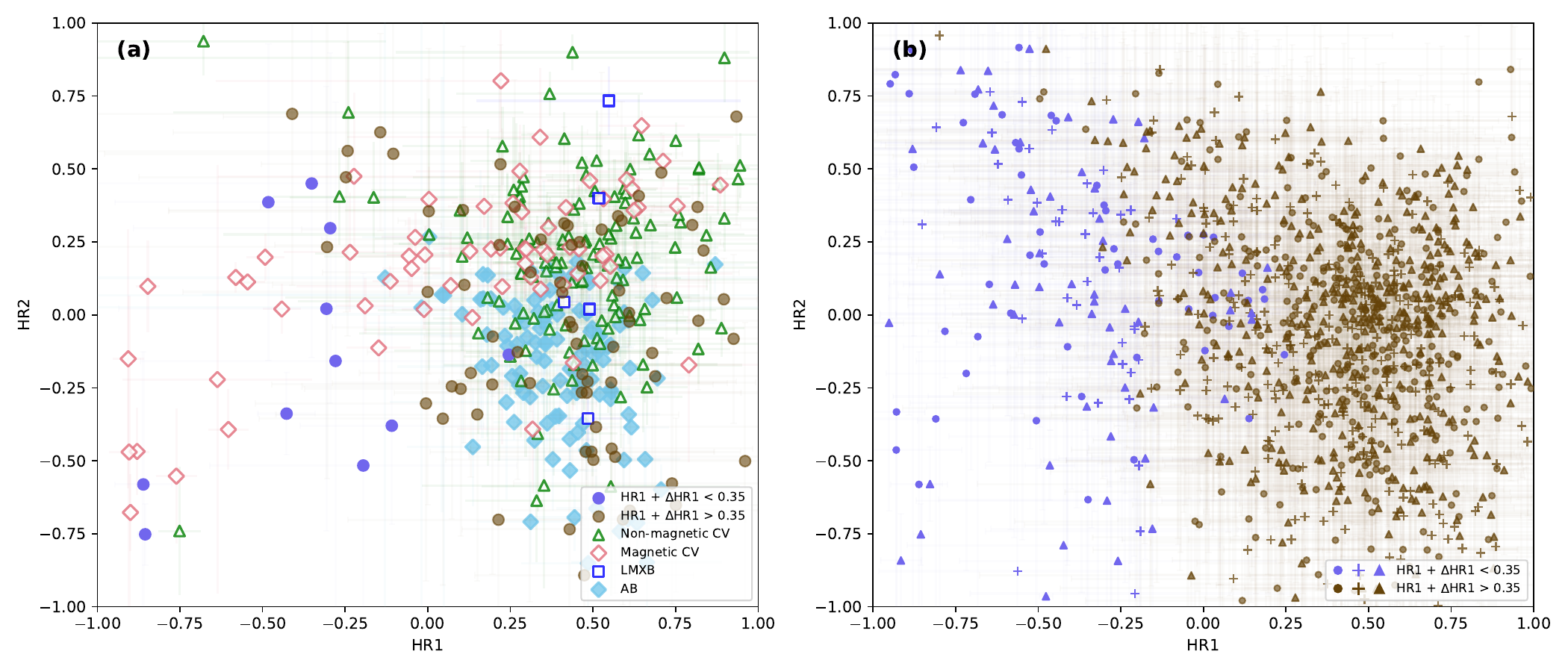} 
\caption{
Panel (a) shows the distribution of HR1 versus HR2 for Sample 1, while panel (b) shows the corresponding distribution for Sample 2. In both panels, the CVs (including both magnetic and non-magnetic CVs), ABs, and LMXBs are from \cite{Ritter2003}, \cite{Eker2008}, and \cite{Fortin2024}, respectively. In panel (a), the purple and brown circles indicate sources satisfying and not satisfying the $\mathrm{HR1}+\Delta\mathrm{HR1}<0.35$ criterion, respectively. In panel (b), the colors have the same meanings as in panel (a), while circles and plus signs indicate sources selected solely by the $d_{\rm L2}>0$ and $L_{\rm X}>10^{31}\ {\rm erg\,s^{-1}}$ criteria, respectively. Sources satisfying both criteria are represented by triangles.
\label{F4}}
\end{figure}

To characterize the expected distributions of different source populations, we selected three comparison samples, including CVs (both magnetic and non-magnetic), LMXBs, and ABs, cross-matched them with eROSITA, and plotted their X-ray properties on the HR1-HR2 diagram. Here, HR1 and HR2 are defined as $\mathrm{HR1}=(C_2-C_1)/(C_2+C_1)$ and $\mathrm{HR2}=(C_3-C_2)/(C_3+C_2)$, where $C_1$, $C_2$, and $C_3$ are the eROSITA counts in the 0.2-0.5, 0.5-1.0, and 1-2 keV bands, respectively. As shown in Figure \ref{F4} (a), magnetic CVs tend to occupy the softer region of the HR1 parameter space, whereas non-magnetic CVs, LMXBs, and ABs largely overlap in the harder region. Following the criterion proposed by \citet{Schwope2024}, 21 out of 65 magnetic CVs (32.3\%) in our comparison sample satisfy $\mathrm{HR1}+\Delta\mathrm{HR1}<0.35$. In contrast, 127 out of 131 non-magnetic CVs (96.9\%) lie at $\mathrm{HR1}+\Delta\mathrm{HR1}>0.35$ and all 112 ABs in our comparison sample also fall into this region.

For Sample 1, 11 sources satisfy the criterion $\mathrm{HR1}+\Delta\mathrm{HR1}<0.35$, suggesting that they may be magnetic CV candidates, as displayed in Figure \ref{F4} (a). However, sources with $\mathrm{HR1}+\Delta\mathrm{HR1}>0.35$ may include non-magnetic CVs, magnetic CVs without a prominent soft X-ray component, as well as ABs and other X-ray source populations.  For Sample 2, as shown in Figure \ref{F4} (b), 156 sources satisfy the criterion $\mathrm{HR1}+\Delta\mathrm{HR1}<0.35$ and may therefore include magnetic CV candidates. However, these results should be interpreted with caution, particularly given that the HR1 measurements for a substantial fraction of sources in both samples are poorly constrained. Overall, the eROSITA HRs provide useful supporting information for identifying potential magnetic CV candidates, but should be regarded as a complementary diagnostic rather than a definitive classification criterion.

\subsection{Cross-Matching with Radio Catalogs}
We further cross-match our Sample 1 (151 sources) and total Sample 2 (1,958 sources) with two major radio continuum surveys: the VLA Sky Survey (VLASS, \cite{Gordon2021}) and the Rapid ASKAP Continuum Survey (RACS,  \cite{Hale2021}) to identify potential radio-bright compact objects. To improve positional accuracy, we adopt the optical counterparts as the reference coordinates, which provide significantly higher astrometric precision than X-ray positions. Cross-matching was performed using a $2^{\prime\prime}$ radius for VLASS and a $5^{\prime\prime}$ radius for RACS, consistent with their respective astrometric uncertainties and beam sizes \citep{Lacy2020,McConnell2020}. We further restrict our radio sample to sources flagged with the morphological classification “S” (single-component sources) to ensure they are point-like, thereby  excluding extended AGN jets, multi-component radio galaxies or unresolved background structures.

\begin{figure*}[ht!]
\centering 
\includegraphics[width=0.7\textwidth]{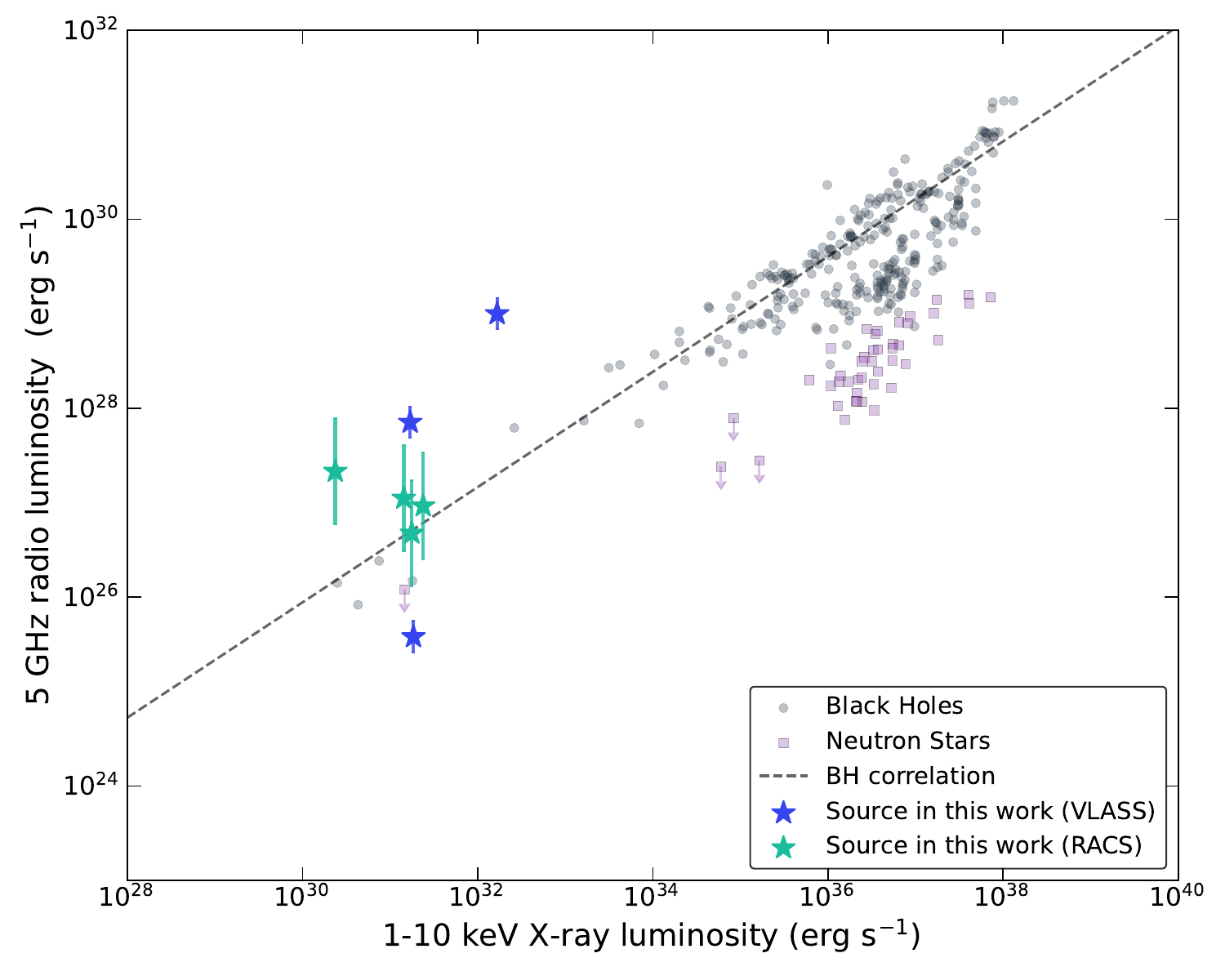} 
\caption{5 GHz radio and 1-10 keV X-ray luminosity of 7 radio sources cross-matched by VLASS and RACS in this work. The Radio/X-ray correlation for accreting compact objects (BH and NS) are from the data base compiled by \cite{bahramian2018}.
\label{F5}}
\end{figure*}

After applying these selection criteria, we obtain a final sample of seven compact radio point sources (see Table \ref{tab2} for details). For each source, we assume a radio spectral index range of $-1.5 \le \alpha \le 0$, encompassing flat to moderately steep spectra consistent with compact synchrotron-emitting sources. Then the radio luminosities were estimated with corresponding error bars by using Gaia-based distance estimates (i.e. $L_{\rm R} = 4\pi d^2 \nu S_\nu$, where $S_\nu \propto \nu^\alpha$ ).

We place these sources on the $L_{\mathrm{X}}$-$L_{\mathrm{R}}$ plane together with known BHs and NSs, as displayed in Figure \ref{F5}. Among the seven radio-detected sources, three sources are found to be consistent with the empirical black hole radio-X-ray correlation within their uncertainties. Besides, 1eRASS J043715.9-471509 is identified as a pulsar in SIMBAD. 1eRASS J040940.7-075329 and 1eRASS J052928.5-585446 are classified in SIMBAD as an RS CVn variable and a CV, respectively. Their relatively low radio luminosities ($L_{\mathrm{R}} \sim 10^{25}\text{-}10^{27}\ \mathrm{erg\ s}^{-1}$) are broadly consistent with their identifications.

For the four remaining sources, the inferred radio luminosities are generally higher than those typically observed in most ABs, although some overlap may still exist with the most radio-luminous AB systems  \citep{Driessen2022,Driessen2024}. Their positions on the $L_{\mathrm{X}}$-$L_{\mathrm{R}}$ diagram are compatible with an XRB interpretation, while MSP and background AGN interpretations cannot yet be excluded. Additional multi-wavelength follow-up observations, particularly improved radio measurements and optical spectroscopic characterization, will be essential to further clarify the nature of these four systems.

\section{CONCLUSIONS AND DISCUSSION} \label{sec4}

In this work, we present a systematic search for COB candidates particularly XRB candidates in the eRASS1 catalog, combining time-domain photometry and multi-wavelength data based on the optical counterpart catalog of \cite{Salvato2025}. We first employ the “X-ray main sequence” as an efficient diagnostic tool to pre-select accreting compact object candidates and to minimize contamination from normal stellar populations. Then, two candidate samples were constructed through complementary pathways to reduce contamination from extragalactic sources such as AGNs: (1) variability-based selection using ZTF time-domain photometry, and (2) distance-based filtering using Gaia astrometric measurements.

Sample 1 contains 151 sources exhibiting reliable periodic variability identified from ZTF time-domain photometric data. By applying a stricter selection criterion (Line 2) to isolate the most promising COB candidates, the sample is further refined to 43 sources. Cross-matching with SIMBAD indicates that these 43 sources include one pulsar, eight CVs, and twelve currently unclassified systems. Sample 2 is constructed using reliable Gaia parallaxes to estimate distances and X-ray luminosities, and contains 1,958 sources with $L_{\mathrm{X}} > 10^{31}\ \mathrm{erg\ s^{-1}}$ or located above Line 2. Cross-matching with deeper X-ray catalogs, including XMM-Newton and Chandra, yields a high-quality subset of 85 sources with improved X-ray characterization.

For both samples, we incorporate ultraviolet data from GALEX and explored the use of $\log(F_{\mathrm{UV}}/F_{\mathrm{X}})$ criterion to distinguish between CVs and LMXBs. Although a sample of 5 XRB candidates is isolated through this method, the reliability of the resulting sample remains limited due to potential contamination from magnetic CVs.  Besides, we further cross-match these two samples with radio surveys including VLASS and RACS, and identified seven radio counterparts. Among these seven sources, one is classified as an RS CVn binaries, one as a CV, and one as a pulsar according to SIMBAD. The remaining four sources lack firm classifications and represent potential XRB candidates, which merit future follow-up observations.

We note that our entire selection process relies on the optical counterpart catalog of \cite{Salvato2025}, which provides reliable associations based on positional matching, photometric properties, colors, morphology, and parallax information. Despite its relatively high reliability, possible misidentifications may still remain. To evaluate the reliability of the adopted optical counterparts in our two samples, we examined the NWAY association quality following the criteria proposed by \cite{Salvato2025}. For the first sample (151 sources), 127 (84.1\%) satisfy \texttt{purity6 > 0.9}, while 136 (90.1\%) satisfy \texttt{p\_any > threshold6}. These fractions in Sample 1 are higher than the corresponding fractions for the overall Galactic-source population (77.9\% and 82.1\%, respectively), suggesting that the optical counterparts in Sample 1 are generally highly reliable. For the second sample (1958 sources),  NWAY association statistics are available
for 1844 sources. Among these sources, 52.5\% satisfy \texttt{purity6 > 0.9}, while 65.5\% satisfy \texttt{p\_any > threshold6}. The lower fractions may result from the inclusion of sources with relatively faint optical counterparts, which can have less secure optical associations and may be selected by the high $F_{\rm X}/F_{\rm opt}$ criterion.   Since our primary goal for Sample 2 is to retain promising XRB candidates, we do not impose a stringent NWAY reliability cut that could preferentially exclude faint optical counterparts and potentially genuine XRBs. We therefore retain optical counterparts with relatively low association reliability, accepting some degree of potential contamination in favor of completeness. Individual systems or small subsamples may still require additional validation of the counterpart associations and source classifications, for example by examining the photometric variability of alternative optical counterparts within the X-ray positional uncertainty regions.

In addition, the current periodicity analysis based on ZTF data may still suffer from aliasing effects and ambiguities in determining the true orbital period. Future higher-cadence time-domain observations, together with complementary surveys such as ATLAS and future facilities like LSST, will help better constrain orbital periods and distinguish between different variability mechanisms. In terms of future multi-wavelength investigations, SED analyses, particularly when combining ultraviolet, optical, and infrared data, may provide additional constraints for distinguishing COBs from normal stellar systems or background AGNs. 

In summary, the combination of eROSITA’s all-sky X-ray data with time-domain data and multi-wavelength surveys provides a powerful and efficient framework for identifying COBs. This approach is readily extendable to other X-ray surveys and will become increasingly powerful with future data releases, offering strong potential for building large and well-characterized samples of COBs. At optical wavelengths, space-based missions such as the Chinese Space Station Telescope (CSST) will significantly expand multi-band photometric coverage. On the X-ray side, future data releases from the eROSITA all-sky survey will further improve statistical completeness and enhance sample sizes.
\begin{acknowledgments}
 We thank the anonymous referee for the insightful suggestions. This work was supported by the National Natural Science Foundation of China under grant 12433007, and the National Key R\&D Program of China under grants 2023YFA1607901 and 2021YFA1600401. The X-ray source catalog used in this work is primarily based on data from eROSITA, the soft X-ray instrument aboard the SRG spacecraft, a joint Russian-German mission supported by the Russian Space Agency (Roskosmos) on behalf of the Russian Academy of Sciences through its Space Research Institute (IKI), and the Deutsches Zentrum f\"ur Luft- und Raumfahrt (DLR). Additional X-ray survey data were obtained from the CSC 2.1 provided by the Chandra X-ray Center (CXC) and from the 4XMM-DR14 catalog compiled by the XMM-Newton Survey Science Centre. We also made use of the LS10 optical counterpart catalog associated with the eRASS1 Main Catalog  (\url{https://erosita.mpe.mpg.de/dr1/AllSkySurveyData_dr1/Catalogues_dr1/}). This research further benefits from data products from the Zwicky Transient Facility (ZTF), the Galaxy Evolution Explorer (GALEX), the Karl G.\ Jansky Very Large Array (VLA), and the Australian Square Kilometre Array Pathfinder (ASKAP).
\end{acknowledgments}

\bibliography{ref}{}

\bibliographystyle{aasjournalv7}

\end{document}